\documentclass[9pt,twocolumn,twoside]{opticajnl}
\journal{opticajournal} 
\usepackage[utf8]{inputenc}
\DeclareUnicodeCharacter{03BB}{$\lambda$}
\DeclareUnicodeCharacter{03C0}{$\pi$}
\setboolean{shortarticle}{false}

\usepackage{subcaption}
\makeatletter
\renewcommand*\subcaption@label{\caption@withoptargs\subcaption@@label}
\makeatother
\usepackage{stfloats}
\usepackage{multirow}
\usepackage{siunitx}
\usepackage{physics}
\usepackage{makecell}
\usepackage{amsmath}
\usepackage{bm}
\DeclareSIUnit\baud{Baud}
\usepackage{caption} 

\usepackage{lineno}

\title{Adaptable Continuous Variable Quantum Network with Finite Size Security}

\author[1,*]{Runjia Zhang}
\author[2]{Akash nag Oruganti}
\author[1]{Huy Q Nguyen}
\author[2]{Ivan Derkach}
\author[1]{Adnan A.E. Hajomer}
\author[2]{Vladyslav C. Usenko}
\author[1]{Ulrik L. Andersen}
\author[1,*]{Tobias Gehring}

\affil[1]{Center for Macroscopic Quantum States (bigQ), Department of Physics, Technical University of Denmark, 2800 Kongens Lyngby, Denmark}
\affil[2]{Department of Optics, Faculty of Science, Palacky University, 17. listopadu 12, 771 46 Olomouc, Czech Republic}
\affil[*]{Corresponding authors: runzh@dtu.dk, tobias.gehring@fysik.dtu.dk}

\begin{abstract}
In recent years, continuous-variable quantum key distribution (CV-QKD) has become a promising paradigm for enabling secure communication among multiple end users sharing the same telecommunication backbone. CV-QKD with reverse reconciliation naturally enables scalability from conventional point-to-point links to quantum access networks based on passive quantum broadcasting channels. Here, we report an experimental demonstration of an active $1:4$ multi-user CV quantum network (QN) in the finite-size regime. With $1.25\cdot10^9$ coherent states exchanged on each $11\text{km}$ quantum channel, the highest performance for secret key generation reach the totals of $1.9\cdot10^{-1}$ bits/channel-use. Furthermore, we investigate adaptable CV-QN protocols that enable network operation across various security and key-rate requirements of individual users. The results establish the practical security of CV-QN compatible with existing telecommunication for broad deployment, and allow an additional degree of freedom for connected end users in existing infrastructures.

\end{abstract}

\setboolean{displaycopyright}{false} 

\begin{document}
  
\maketitle

\section{Introduction}


Quantum key distribution (QKD) as the most commonly developed application in quantum communication is progressing towards full-scale implementations~\cite{usenko2026continuous}, providing information-theoretical security built on the principles of quantum physics to secure information infrastructure as the projected timeline for a cryptographically relevant quantum computer has compressed significantly~\cite{cain2026shor}. QKD protocols allow two distant parties to generate two identical bit-strings decoupled from potential eavesdropper via an authenticated otherwise insecure public channel. Such ideal keys can be used as symmetric keys for further cryptographic tasks to secure information transmission over a shared communication infrastructure. Contemporary information infrastructure orients around multi-user communication, accommodating the distanced exchange of information among many users. In recent years, multi-user continuous-variable (CV-QKD) has become a nascent but promising alternative to this so-called last-mile challenge connecting end users to the same telecommunication backbone~\cite{hajomer2024continuous,qi2024experimental,bian2023high,liu2024integrated}.

Point-to-point CV-QKD has shown substantial potential of full-scale telecommunication integration in recent years. CV-QKD has demonstrated remarkable performance with respect to secret key rate and transmission distance ~\cite{Huang2016LongDistance, HuangLin2015_1Mbps, Zhang2020LongDistance, huang2015high, leverrier2010cvqkd, hajomer2025coexistence}. Furthermore, CV-QKD has shown compatibility with telecommunication infrastructure in that quantum correlations can be distributed with information encoded and decoded in the phase and amplitude quadratures of the optical field using standard telecommunication compatible components in coherent optics as well as coexist with classical information transmission in the same fiber~\cite{Jouguet2012Analysis,qi2010feasibility, hajomer2025coexistence, jouguet2013experimental, kumar2015coexistence}. Complete protocol implementation with security analysis on imperfect devices with realistic assumptions has been meticulously demonstrated on simple optical platforms enabled by digital signal processing (DSP) module to achieve composable finite-size security against collective attacks~\cite{jain2022Practical, chin2022digital, kanitschar2023finite, hajomer2025experimental_QPSK}. All of the above establish that point-to-point CV-QKD's feasibility in large-scale deployment for practical security.  

As CV-QKD utilizes the wave property of light, a point-to-point CV-QKD system can be almost seamlessly extended to a point-to-multi-point network connecting parallel end users. With the addition of a passive optical beam splitter, the coherent states from Alice are split into different modes, each traveling through a subsequent quantum channel before being measured by each Bob. 
After the quantum phase of establishing correlation, Alice reconciles the measurement results with all Bobs individually. The CV quantum network (CV-QN) enables simultaneous key generation with resource, which is correlation, for each user. This differs from the previous approach with discrete-variable (DV) systems which rely on probabilistic or time-sharing strategy~\cite{townsend1997quantum, frohlich2013quantum}. Such a CV multi-user solution is desirable for the quantum access network to approach the total secret key volume and generation speed, network capacity, composable finite-size security, as well as the adaptability requirements of current and future communication infrastructure ~\cite{hajomer2024continuous,bian2026approaching}.


Despite recent reports of successful experimental implementations of CV quantum networks with 4, 8, and 16 users on various platforms \cite{huang2021realizing, bian2023high, liu2024integrated, pan2025high, hajomer2024continuous, bian2026approaching},
two key challenges call for further investigation. The first regards the fact that the asymptotic limit is an ideal assumption that overestimates secret key rate which undermines security. Simultaneous finite-size key generation among all users in the network has not yet been experimentally realized. 
The second open frontier is with respect to the network performance. 
Adapting point-to-point CV-QKD protocols to network settings (referred to as untrusted broadcast protocol), where users operate independently, leads to poor scalability and limited total key volume, as inter-user correlations are neither exploited as a resource nor properly accounted for \cite{oruganti2025multiuser,hajomer2024continuous,bian2026approaching}. In the untrusted broadcast protocol \cite{huang2021realizing,hajomer2024continuous}, each user treats the fraction of the signal received by others as fully accessible to Eve, yielding keys that are independent of both Eve and all remaining users (at the cost of a conservative upper bound on Eve's information). However, recent work has shown that introducing trust assumptions or inter-user collaboration significantly improves both per-user and joint key rates \cite{hajomer2024continuous,bian2026approaching,takeoka2017unconstrained}, motivating a systematic study of how inter-user correlations can be consistently and optimally attributed across the full multi-user network.



Here, we present a comprehensive security analysis of multi-user CV-QKD networks across all relevant trust scenarios among the network participants. Rather than focusing on a single operational regime, we adopt a unified framework that allows Alice to flexibly assign the role of the non-reference users according to the required security setting, namely as trusted, collaborative, or untrusted parties. This holistic approach provides a consistent way to analyze how inter-user correlations can be either exploited or excluded depending on the protocol objective.

Unlike approaches that construct per-user end-to-end key rates and sum them towards a collective upper benchmark \cite{bian2026approaching}, we instead begin from the collective upper bound on the joint key and decompose it exactly into successive, independent per-user contributions ~\cite{oruganti2025multiuser}. The sum of these individual contributions is therefore not an approximation of the joint key rate, but is analytically equal to it by construction. 
Together, these results pave the way for scalable, adaptable CV-QN with with dynamic resource allocation that deliver flexibly adjusted and high secret key rates under finite-size constraints across the full range of trust scenarios.


\section{Adaptable quantum network protocols}
\label{sec: protocol}

\subsection{General network and security framework}

\begin{figure*}[ht]
    \centering
    \includegraphics[width=0.99\linewidth]{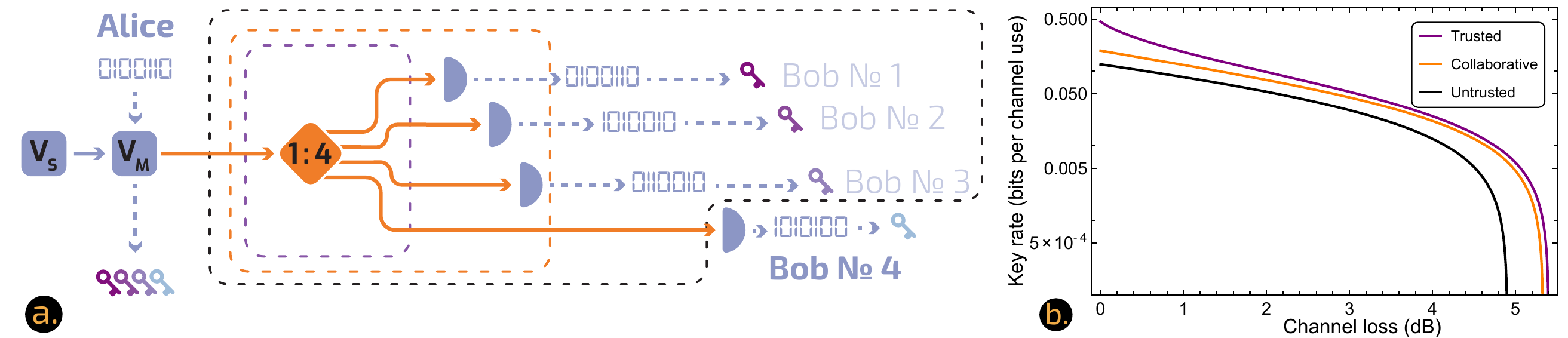}
    \caption{
(a) Trust assumptions of different broadcasting CV-QKD protocols in access CV-QKD network. Dashed areas indicate elements that are \textit{not trusted} under respective protocol. Under the trusted protocol quantum channels are not trusted, while users have a varying degree of trust between them. The collaborative protocol relies on measured results of other users to be kept private from Eve, while their detectors are not necessarily trusted. All non-reference Bobs are assumed to be fully a part of Eve in the untrusted protocol. (b) Per-user secret key rate as a function of channel loss for a multi-user CV-QKD network with uniform splitting among users. 
The solid curves correspond to the maximal per-user key rate that can be achieved with  untrusted, collaborative, and trusted protocols. 
The modulation variance is fixed to \(V_M = 5~\mathrm{SNU}\), with reconciliation efficiency \(\beta = 0.95\). 
The channel excess noise is fixed to \(5~\mathrm{mSNU}\). 
The trusted receiver parameters are \(\eta_{\mathrm{d}} = 0.68\) and \(\nu_{\mathrm{el}} = 60~\mathrm{mSNU}\). 
All key rates are expressed in bits per channel use.
    }
    \label{fig:per_user_keyrate_trust_scenarios}
\end{figure*}
\label{subsec: general network and security framework}
A QKD-based network is a complex communication system consisting of interconnected nodes, quantum and classical channels, that can implement various QKD protocols enabling an information-theoretically secure key exchange between multiple participants. Although there are numerous network configurations available, we focus on a common telecommunication technology for broadcasting data (network access) to end-points using point-to-multipoint topology, known as Passive Optical Network (PON), as shown in  Fig.~\ref{fig:per_user_keyrate_trust_scenarios} (a).

Network operates over the series of rounds where Alice generates a sequence of quantum signal states and modulates them (with a variance of Gaussian modulation distribution given by $V_M$), then sends them via quantum channel to $M$ Bobs (in the following we use terms ``Bob'' and ``user'' interchangeably), that each performs a coherent detection of the signal. After this stage, Alice holds her modulation data, while the users hold correlated measurement data. Quantum channel is comprised of a fiber link prior to a $1:M$ beam splitter, and a series of $M$ links connected to users. Each link is characterized by a transmittance $\eta_m$ and excess noise $\epsilon_m$. Every Bob then individually uses an authenticated classical channel to perform error-correction, and parameter estimation with Alice, which, after data post-processing, can reconstruct the overall covariance matrix and proceed to security analysis of the network. We work within the Gaussian framework, for which the security analysis under collective attacks is fully determined by the covariance matrix of the joint state shared by Alice and all users \cite{weedbrook2012gaussian}. 
In this sense, the covariance matrix of the global network state is the fundamental object from which all relevant security quantities are obtained. 
Users conduct parameter-estimation procedure independently and it is sufficient to employ standard techniques used in point-to-point Gaussian-modulated CV-QKD protocols~\cite{leverrier2010finitesize,Jouguet2012Analysis}. 
We take into account finite data block lengths and determine confidence interval of all channel parameters $\eta_m$ and $\epsilon_m$ where $m\in (1, M$), and minimize the key rate by choosing values within respective confidence intervals that would result in a lower bound on the secure key rate, which is is used as a benchmark for the network performance. Generally, the key rate between Alice and a specific user \(B_k\) is defined as: 

\begin{equation}\label{eq:basic-key}
    K_k^{\mathrm{FS}}
    =
    \beta I_{AB_k}
    -
    \chi_{B_kE}^{\varepsilon_{\mathrm{PE}}}
    -
    \Delta(N),
\end{equation}
where \(\beta\) is the reconciliation efficiency, \(I_{AB_k}\) is the classical mutual information between Alice and \(B_k\) that can be directly assesed from the measurement results, and \(\chi_{B_kE}^{\varepsilon_{\mathrm{PE}}}\) is the Holevo information evaluated using the covariance matrix obtained after post-processing. The term \(\Delta(N)\) denotes the standard finite-size correction used in composable CV-QKD security analysis~\cite{leverrier2010finitesize,hajomer2025coexistence}. The approach to evaluation of the rate (\ref{eq:basic-key}) is determined by the specific protocol Alice and Bobs have pre-agreed to employ.

\subsection{Multi-user protocols}
\label{sec:protocols}

A protocol is a set of rules governing how the raw measurement data collected from a fixed experimental setup is processed, reconciled, and analyzed to extract a secure key — with different protocols applying distinct post-processing strategies and security assumptions, and consequently yielding different key rates from the same underlying hardware. In other words, the same network data can also be used in different operational modes, depending on which users are trusted, which users participate in reconciliation, and whether the goal is joint key generation or maximising the key rate of a selected user. We consider three distinct CV-QKD multi-user protocols:  \textit{untrusted, collaborative} and \textit{trusted} protocols.\\
\textit{In the untrusted protocol} every user distills the key with Alice independently under a conservative assumption that others (i.e. non-reference Bobs) are potential adversaries, or more specifically part of the joint state of the eavesdropper Eve.  
For a given user $B_k$, both the mutual information and the Holevo bound are evaluated from the reduced covariance matrix $\Gamma_{AB_k}$. Operationally, this attributes all correlations involving the other users to Eve, hence leads to the lowest resulting key between each pair $A-B_k$. Such protocols has identical performance as standard point-to-point CV-QKD protocols \cite{huang2021realizing, hajomer2024continuous,usenko2026continuous}. 

\textit{In the trusted protocol}, while a single user $B_k$ generates a key with Alice, some of the other users can be assumed to be trusted, i.e. their signals were faithfully received by respective users, with measurement results not disclosed to Eve. Alice establishes a hierarchical trust ordering: for each reference user $B_k$, a subset of the remaining users is classified as trusted, with the size of this subset increasing successively across users. This hierarchy enables all users to extract independent keys from a single round of measurements, with each successive user benefiting from a larger trusted set and thus achieving a progressively higher key rate \cite{hajomer2024continuous}. For a user $B_k$ who trusts all other users, the mutual information can be evaluated from the reduced covariance matrix $\Gamma_{AB_k}$, since it only concerns Alice and the reference user. However, the Holevo bound is evaluated from the global covariance matrix $\Gamma_{AB_1\cdots B_M}$, because the remaining trusted users are excluded from Eve's purification system. This allows for the selected user to reach the highest single-user key rate (\ref{eq:basic-key}), and provides a significant key improvement over the untrusted protocol \cite{takeoka2017unconstrained}. 

\textit{In the collaborative protocol}, each user assumes, similarly as in the trusted protocol, other users faithfully received broadcasted signal (crucially, without any assumption of trust on their detector noise), however that the measurement results were disclosed to Eve. The relevant quantum state is then the conditional state of Alice and $B_k$, obtained by conditioning on the disclosed outcomes of all assisting users. This conditioning tightens the estimation of Eve's accessible information: rather than conservatively assigning the full correlations of the remaining users to Eve (as in the untrusted protocol), their disclosed data directly constrains Eve's knowledge \cite{bian2026approaching}. Consequently, both the mutual information $I(A{:}B_k)$ and the Holevo bound $\chi(B_k{:}E)$ are evaluated from the conditional covariance matrix, which encodes the residual Alice–$B_k$ correlations after the public supposed measurement results announcements of the other users have been taken into account:

\begin{equation}
    \Gamma_{AB_k \mid \bm{y}_{\neg k}},
\end{equation}
where $\bm{y}_{\neg k}$ denotes the collection of measurement outcomes disclosed by all users other than $B_k$. In this scenario the mutual information is correspondingly conditioned on the disclosed data,
\begin{equation}
    I_{AB_k}^{\mathrm{collab}}
    \equiv
    I(A:B_k \mid \bm{y}_{\neg k}).
\end{equation} 
This protocol outperforms the untrusted protocol, however, due to stricter security assumptions, yields lower per-user key rate than can be achieved under the trusted protocol, as shown in Fig.~\ref{fig:per_user_keyrate_trust_scenarios}(b).

\subsection{Network performance bound}
\label{sec:multi_user_workflow}


\begin{figure*}[ht]
    \centering
    \includegraphics[width=0.95\textwidth]{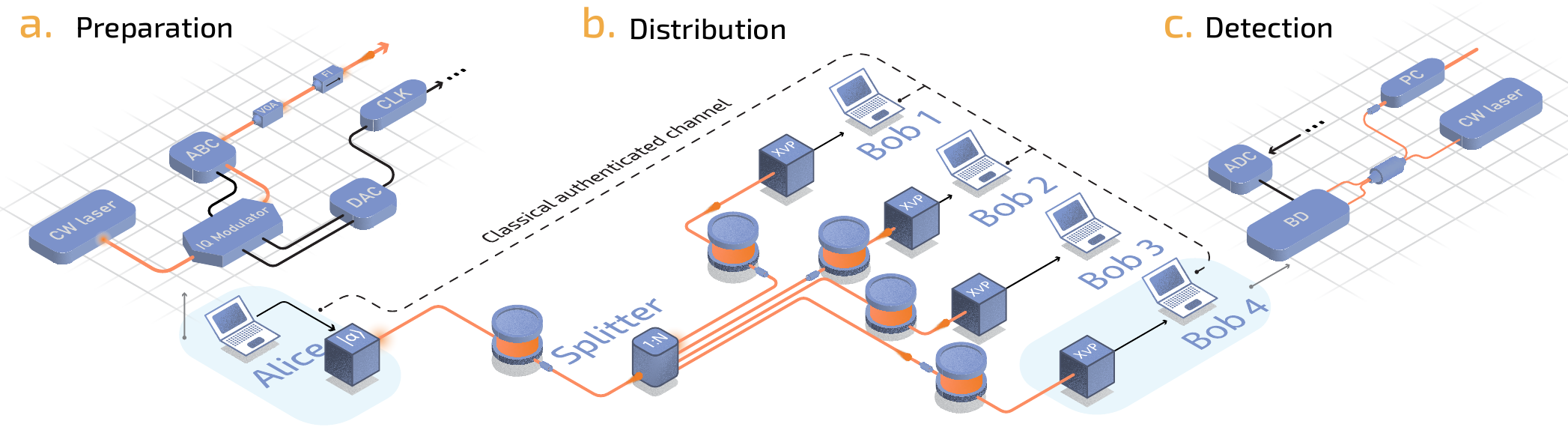}
    \caption{ A continuous variable quantum access network in action in the lab. (a) State preparation station comprises a signal laser, a digital-to-analog converter (DAC), an IQ modulator stabilized by an automatic bias controller (ABC), as well as a variable optical attenuator (VOA) and an isolator. (b) State distribution is assisted with a $1:4$ splitter which splits the coherent states, forming a quantum broadcasting channel that connects N users while distributing quantum correlations evenly among them. The link consists of two segment, a 10 km fiber spool before the $1:4$ splitter represent the communication backbone and at each output of the splitter represent last-mile connection to the end-users. The quantum state sent by Alice is mixed with different quantum noises, consequently. (c) A receiver station comprises a balanced detector (BD) that receives both the signal and the local oscillator (LO) at the 50:50 beamsplitter (BS), while each resultant signal is detected by a photodetector. The analog readout is converted to digital by the analog-to-digital converter (ADC).}
    \label{fig: the networked system}
\end{figure*}

The distinction between the aforementioned multi-user protocols arises in the classical post-processing stage and in the security interpretation of the users' data. 
If reconciliation succeeds in the network, Alice and the users can bound the secret key extractable from the joint correlations shared between Alice and \(\mathbf B=(B_1,\ldots,B_M)\). In the Gaussian framework, this network-wide key content is determined by the covariance matrix of Alice and all users. The corresponding finite-size joint key rate can be written as
\begin{equation}
    K_{\rm joint}^{\rm FS}
    =
    \beta I(A:B_1,\ldots,B_M)
    -
    \chi^{\epsilon_{\rm PE}}(B_1,\ldots,B_M:E)
    -
    M\Delta(N).
    \label{eq:joint_finite_size_key}
\end{equation}
This quantity upper bounds the total secret key content available in the network (between Alice and the complete user system), and is conceptually distinct from individual user key rates in Eq.~\ref{eq:basic-key}. Earlier works on CV-QKD access networks have proposed user-wise key-rate expressions for point-to-multipoint settings, each resting on different assumptions regarding the role of non-reference users, the trust model, and the network architecture \cite{hajomer2024continuous,bian2026approaching,pan2025high}. While such expressions are valuable for comparing network operation modes, they do not automatically guarantee a consistent or optimal attribution of the key content residing in the full multi-user state. In particular, evaluating Alice–$B_k$ key rates independently risks double-counting correlations shared across multiple users, with the consequence that the sum of individual rates may exceed the key rate extractable from the joint Alice–$\mathbf{B}$ state.

To avoid this overcounting, we use the chain-rule decomposition introduced in Ref.~\cite{oruganti2025multiuser}. The joint mutual information can be expanded by adding the users one by one in a chosen order,
\begin{align}
    I(A:B_1,\ldots,B_M)
    &=
    I(A:B_1)
    +
    I(A:B_2|B_1)
    +\\
    &\cdots 
    +
    I(A:B_M|B_1,\ldots,B_{M-1})
    \nonumber \\
    &=
    \sum_{k=1}^{M}
    I(A:B_k|B_1,\ldots,B_{k-1}).
    \label{eq:chain_rule_mutual_information}
\end{align}
Each term gives the additional correlation with Alice contributed by user \(B_k\), conditioned on the users that have already been included in the chosen ordering.

The Holevo contribution can be decomposed in the corresponding telescopic form,
\begin{align}
    &S(\rho_{AB_1\cdots B_M}) - S(\rho_{A|B_1\cdots B_M})
    \nonumber \\
    &=
    \left[
        S(\rho_A)-S(\rho_{A|B_1})
    \right]
    +
    \left[
        S(\rho_{A|B_1})-S(\rho_{A|B_1B_2})
    \right]
    + \cdots
    \nonumber \\
    &\quad
    +
    \left[
        S(\rho_{A|B_1\cdots B_{M-1}})
        -
        S(\rho_{A|B_1\cdots B_M})
    \right]
    \nonumber \\
    &=
    \sum_{k=1}^{M}
    \left[
        S(\rho_{A|B_1\cdots B_{k-1}})
        -
        S(\rho_{A|B_1\cdots B_k})
    \right].
    \label{eq:telescopic_holevo_decomposition}
\end{align}
All intermediate conditional entropies cancel pairwise, leaving the Holevo contribution associated with the full joint state.

Combining~\eqref{eq:chain_rule_mutual_information} and \eqref{eq:telescopic_holevo_decomposition}, the joint key rate can be written as a sum of individual assigned contributions,
\begin{equation}
    K_{\rm joint}^{\rm FS}
    =
    \sum_{k=1}^{M}
    K_k^{\rm dec},
    \label{eq:joint_key_decomposition}
\end{equation}
here \(K_k^{\rm dec}\) is the incremental contribution assigned to user \(B_k\), conditioned on the users that precede it in the chosen ordering. The finite-size correction can either be applied at the level of the total joint key rate, as in~\eqref{eq:joint_finite_size_key}, or distributed consistently among the assigned user contributions. In either case, the defining consistency requirement is that the assigned individual contributions sum to the joint finite-size key rate.

This decomposition is optimal in the specific sense that it exactly exhausts the available joint key content without reusing correlations shared among several users. The individual contributions may depend on the chosen ordering, but the sum is invariant and equals the joint key rate. The ordering should therefore be understood as an attribution rule, not as a required temporal order of reconciliation. In practice, the users may send reconciliation information to Alice in parallel, after which the network-wide key content can be decomposed into individual contributions according to the chosen ordering.

\section{A CV quantum network implementation}

The proof-of-concept CV-QN is shown in Fig.~\ref{fig: the networked system}. The system, supporting four active end users, consists of an optical and a digital module, both optimized to ensure finite-size security. In the optical module, the LO was tuned to maintain sufficient clearance above electronic noise without inducing detector instability. In the digital domain, the detector gain was calibrated relative to the LO power. The coupled dynamic variables were optimized together. Additionally, the pilot tone power was adjusted to provide robust phase recovery without saturating the detectors. Since the setup involves four receiver systems, these optimizations were executed iteratively and systematically until the high stability and low noise required were achieved to generate finite-size keys for all four users. To establish quantum correlations, Alice distributed coherent states among the Bobs, encoding information using symbols sampled from a Gaussian distribution. Each symbol's quadrature components, $x_i$ and $p_i$, were independent and identically distributed (i.i.d.) random variables generated by a quantum random number generator (QRNG). Symbols were transmitted at a rate of 125 MBaud and upsampled to 8 samples per symbol using a root-raised cosine (RRC) pulse filter with a roll-off factor of 0.2. Adopting single-sideband modulation, the quantum signal was frequency-shifted by $150$ MHz relative to the carrier and a pilot tone at 10 MHz was frequency-multiplexed to assist the digital signal processing (DSP) process. A 1 GSample/s 16 bits resolution converted the digital samples to an electrical analog signal to drive the two arms of an IQ modulator. Around $1.25\cdot10^9$ symbols were exchanged between Alice and each Bob. The power of the optical signal was controlled by a programmable electrical VOA and sent through a $10$km single-mode fiber channel. Subsequently, the signal was split into four by a passive 1:4 beamsplitter, followed by an additional 1 km fiber channel before entering Bob's receiver stations. Each Bob received the symbols through a distinctive channel corrupted by noise from the channel assumed to be controlled by Eve.





Each Bob detected the signal coming from the channel using radio-frequency (RF) heterodyne detection with a locally generated local oscillator (LLO) at frequency $f_{LO}$ interfering with the signal with the carrier $f_s$ on a 50:50 beamsplitter resulting in a beating frequency $f_b = \abs{f_{LO} - f_s}$ within the electronic detection bandwidth. Each user had a polarization controller to match the polarization of the incoming signal with that of the LLO to maximize the visibility. The signal were detected using balanced detector (BDs) and sampled by a digital-to-analog converter with the sampling rate of $1$Gsample/s. 

DSP was performed on the received signal to recover the quantum symbols~\cite{chinMachine2021,hajomerLongdistance2024}. A whitening filter was applied to remove time-dependent correlations from the detector’s transfer function. Utilizing the strong pilot signal, we recover the carrier frequency and phase using a machine learning-based technique~\cite{chinMachine2021}. Subsequently, the propagation delay was compensated for, and the signal was downsampled using a matched filter.

\section{Result}
\subsection{Number of symbols exchanged}
\begin{figure*}[ht]
    \centering
    
    \includegraphics[width=0.32\textwidth]{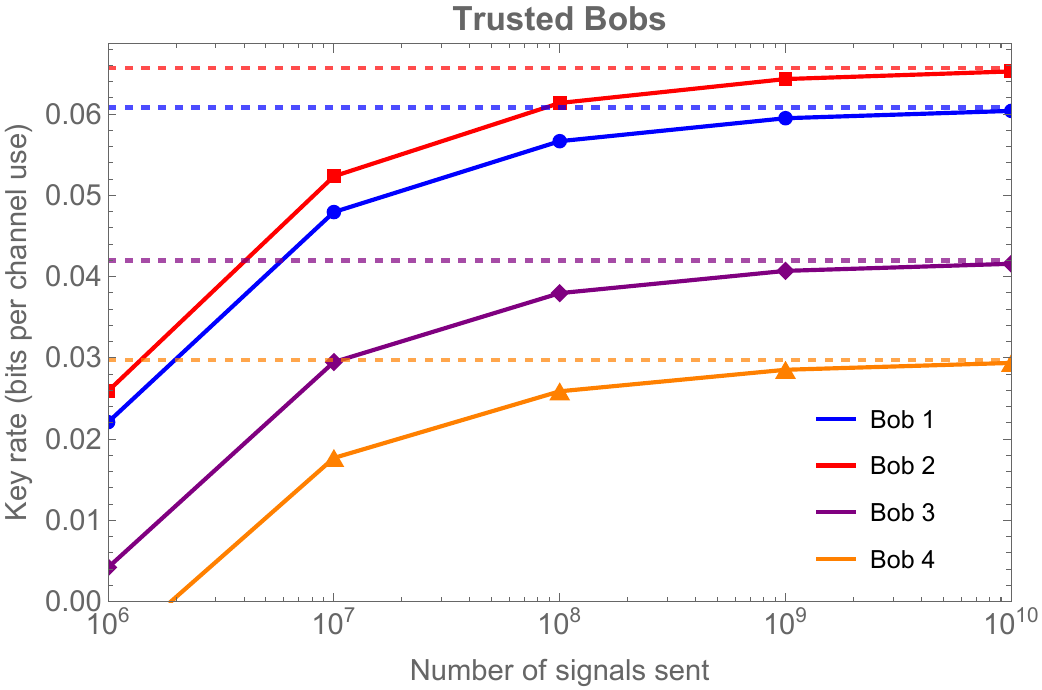}
    \hfill
    \includegraphics[width=0.32\textwidth]{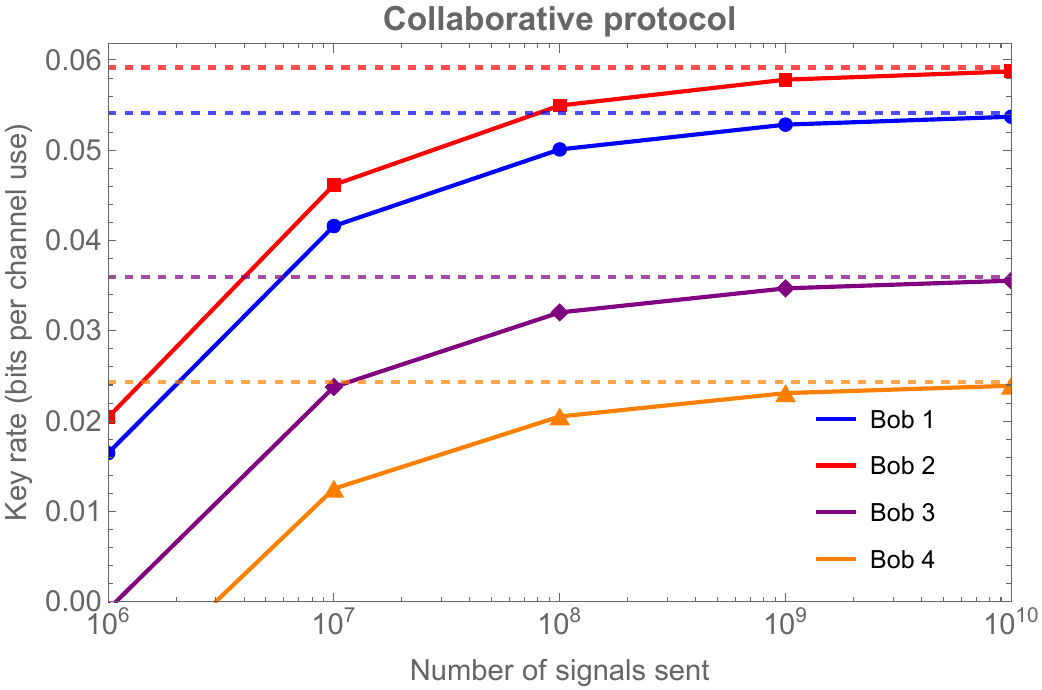}
    \hfill
    \includegraphics[width=0.32\textwidth]{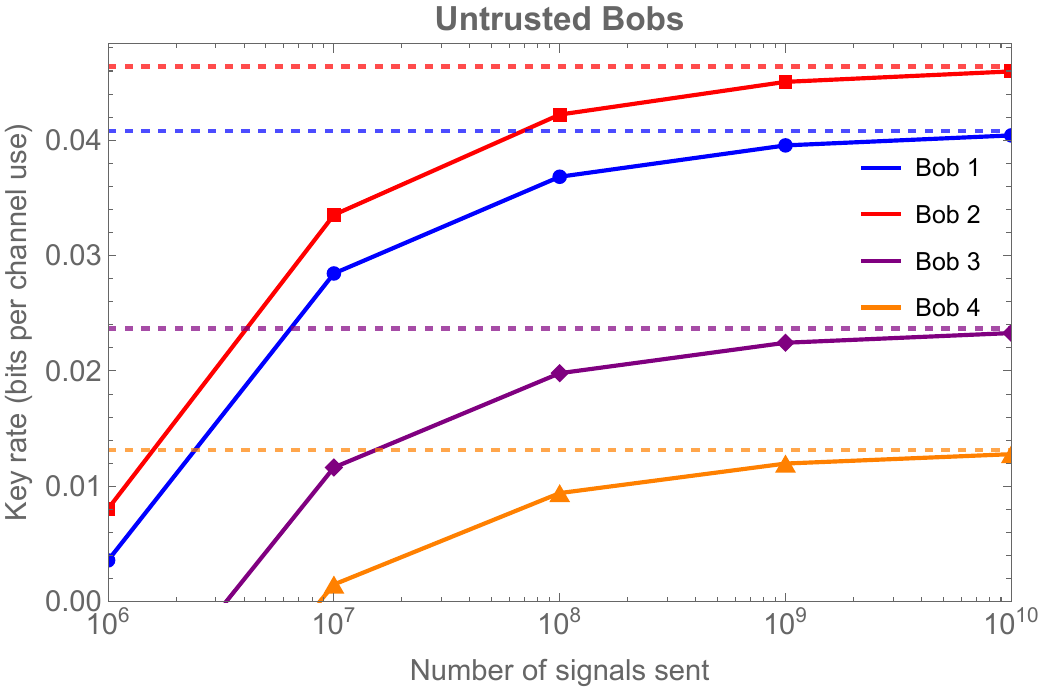}
    
    \caption{
    Finite-size secret key rates as a function of the total number of exchanged quantum signals for the different trust scenarios considered in the multi-user CV-QKD network. 
    The solid curves correspond to the finite-size secret key rates, while the dashed horizontal lines indicate the corresponding asymptotic secret key rates.  (Left) Trusted-user scenario, where non-reference users are assumed trusted and excluded from Eve's system. 
    (Middle) Collaborative (conditionally trusted) scenario, where auxiliary users publicly disclose their measurements to improve the effective correlations used for key extraction. 
    (Right) Fully untrusted-user scenario, where all non-reference users are treated as part of Eve's purification. }
    
    \label{fig:finitesize_trust_comparison}
\end{figure*}

The networked system can operate stably for long measurements, which is ideal for experimentally demonstrating multi-user protocols in the finite-size regime where several statistical penalties necessitate a larger number of symbols exchanged. Fig.~\ref{fig:finitesize_trust_comparison} shows finite-size secret key rates with respect to number of symbols $N = 10^x$, $x\in[6,10]$ with respect to each user under \textit{trusted}, \textit{collaborative} and \textit{untrusted} protocols. With $N = 1.25\cdot 10^9$ as this experiment demonstrated, we operate close to the asymptotic limit indicated by horizontal dashed lines.  

\subsection{System calibration and parameter estimation on the networked system} 
\label{sec:PE}


The quantum phase consisted of multiple rounds of preparation, distribution, and detection. The parameters estimated from the measurements of each Bobs are summarized in Table ~\ref{table: 1.25e9_calibration and parameter estimation}. This followed by classical phase consisting of information reconciliation and privacy amplification. Alice generated $1.25\cdot10^9$ coherent states with a modulation variance of $V_M =5.04$ SNU. As Eve has no access to any Bob's detection station, electronic noises of the detectors are considered as trusted, and the detector efficiency of $\tau =0.68$ is taken as trusted loss. The excess noise is kept relatively low through fine-tuning the system's parameters. The deviation from the channel loss of the users in comparison with the intrinsic ~8.2 dB loss can be explained by the imperfection in the passive beamsplitter and polarization mode-mismatch.  It is worth noting that, although a common state is distributed, the measurement outcomes remain unique since each Bob's mode is coupled with independent vacua. Using reverse reconciliation, Alice can establish distinct and independent keys with each Bob.

\begin{table}[ht]
    \centering
      \caption{Calibration for the system: $\eta$: untrusted transmittance, trusted noise for $1.25\cdot10^9$ symbols, modulation variance $V_M$: $5.04$ SNU. Parameter estimation with $1.25\cdot10^9$ symbols }
    \begin{tabular}{c c c c}
    \toprule
        User & \makecell{Trusted noise \\ (mSNU)}  & $\eta$ & \makecell{Excess noise \\ (mSNU)} \\
    \midrule
        1 & 54.00 & 0.13 (-8.94 dB) & I: 4.24, Q: 4.10\\
        
        2 & 49.80 &  0.12 (-9.23 dB) & I: 2.94, Q: 2.98\\
        
        3 &  60.22 & 0.11 (-9.42 dB) &  I: 5.02, Q: 5.00\\

        4 & 51.08 & 0.10 (-10.02 dB) & I: 5.14, Q: 5.18\\
        
  \bottomrule
  \end{tabular}
  \label{table: 1.25e9_calibration and parameter estimation}
\end{table}

Table~\ref{tab:finite_size_keyrates_models} presents the finite-size secret key rates for each user under different trust protocols. As expected from the theoretical analysis in Fig.~\ref{fig:per_user_keyrate_trust_scenarios}(b), the trusted protocol yields the highest key rate, with User 2 achieving 0.0644 bits/channel use and the total key rate in the network reaching 1.9 bits/channel use. In the most pessimistic scenario, the total key rate is reduced to 0.1192 bits/channel use. Interestingly, the collaborative protocol does not sacrifice significant network performance, maintaining a maximum individual key rate of 0.0579 bits/channel use and a total key rate of 0.1688 bits/channel use.

\begin{table}[ht]
\centering
\caption{Finite-size secret key rates ($K_i$ in bits/channel use) for different trust models with $\beta=0.95$ and block sizes $N=1.25\cdot10^9$.}
\label{tab:finite_size_keyrates_models}
\begin{tabular}{c c c c}
\toprule
 User & $K_i^{\mathrm{(untrusted)}}$ & $K_i^{\mathrm{(collaborative)}}$ & $K_i^{\mathrm{(trusted)}}$ \\
\midrule
 1 & 0.0396 &  0.0529 & 0.0596 \\
 2 & 0.0451 & 0.0579 &  0.0644 \\
 3 & 0.0225 & 0.0348 &  0.0408 \\
4 & 0.0120 & 0.0232 &  0.0286 \\
\bottomrule
\end{tabular}
\end{table}

\begin{table*}[!htp]
\label{table: quantitative comparison}
\centering
\caption{Quantitative comparison with other reported Gaussian modulated 1-to-N CV network against collective attack. We take the maximum total SKR totaling all users supported by the network, according to respective multi-user protocol in the literature as a figure of merit. The reasoning behind is that this figure is achieved for independent key generation.}
\label{tab: conclusion}
\begin{tabular}{| c | c | c | c | c | c |}
\toprule
 {Work} & \makecell{1-to-N \\ (Active Users)} & {Distances~(km)} & \makecell{Sum of SKR \\ (bit/channel use)} & Finite-size  &  Requirement  \\
\midrule
 this work & $1:4$~(4) & $10 + 1$ & $1.9\times10^{-1}$& Yes & successful information reconciliation for all Bobs\\

 ~\cite{bian2026approaching} & $1:2$~(2) & $25 + 5$ &   $0.89\times10^{-2}$ & No &  collaborative protocol\\ 
 
 ~\cite{hajomer2024continuous} & $1:8$~(8) & $10 + 1$ & $3.1\times10^{-2}$ & No &  trust protocol \\
 
 ~\cite{pan2025high} & $1:16$~(4) & $5 + 1$ &
 $2\times{10^{-3}}$ & No & trust protocol\\
 
\bottomrule
\end{tabular}
\end{table*}

\section{Conclusion}

As point-to-point QKD systems reach technical maturity and enter the commercialization phase, the need to scale quantum network infrastructure has become critical. CV-QKD excels in scalability because it integrates seamlessly with current telecom hardware and supports simultaneous multi-user key generation \cite{hajomer2025coexistence,hajomer2024continuous,bian2023high}. Here, we present a comprehensive finite-size security analysis and experimental demonstration of a multi-user CV-QKD network operating under multiple trust scenarios. Rather than restricting the study to a single operational regime, we analyze \textit{trusted}, \textit{collaborative}, and fully \textit{untrusted} protocols within a unified Gaussian security framework. The results illustrate how the achievable secret key rates depend fundamentally on the security assumptions imposed on the network participants and conditional of information reconciliation. In particular, we demonstrated how the same underlying physical network can support qualitatively different operational modes and resource-allocation strategies depending on whether inter-user correlations are excluded, publicly disclosed, or fully exploited. This can be used to substantially improve both individual and total secret key rates compared to the untrusted broadcast approach. In real-world networks, user security and communication demands are rarely static, making elasticity a vital property. Our comprehensive analysis serves as the foundation for a flexible quantum network, one that is dynamically adaptable to involve security concerns and bandwidth requirements.


More importantly, this work establishes an analytical framework for reaching the fundamental joint key rate limit of the network. Unlike previous approaches that construct independent per-user rates whose sum merely approaches the collective upper bound~\cite{bian2026approaching}, our method begins directly from the network joint key rate and decomposes it exactly into successive independent user contributions through a chain-rule decomposition of the mutual information and Holevo terms. As a result, the sum of the assigned user contributions is analytically equal to the total joint key content by construction, avoiding the double counting of correlations shared among multiple users. This demonstrates not only how to approach the network key-rate limit, but how to consistently and optimally attain it within the Gaussian finite-size framework.

These results provide both a practical and theoretical foundation for adaptable quantum access networks with dynamic resource allocation, elastic trust management, and optimal utilization of shared quantum correlations. We expect this framework to play an important role in the development of future large-scale quantum communication infrastructures and multi-user quantum cryptographic services.


\begin{backmatter}

\bmsection{Acknowledgments} 
RZ, HN, AAEH, ULA and TG acknowledge support from the Danish National Research Foundation, Center for Macroscopic Quantum States (bigQ, DNRF142). This project was funded within the QuantERA II Programme (project CVSTAR) that has received funding from the European Union’s Horizon 2020 research and innovation programme under Grant Agreement No 101017733. ID acknowledges support from the project 22-28254O of the Czech Science Foundation. AO and VCU acknowledge support from the projects 21-44815L of the Czech Science Foundation and 8C22002 of the Czech Ministry of Education, Youth and Sports (MEYS), VCU acknowledges the project CZ.02.01.01/00/22\_008/0004649 (QUEENTEC) of MEYS. We acknowledge support from  European Union’s Horizon Europe research and innovation programme under the project ``Quantum Security Networks Partnership'' (QSNP, grant agreement no. 101114043), from the European Union’s Digital Europe programme (QCI.DK, grant agreement no. 101091659 and BeQCI, grant agreement no. 101091625), and from Innovation Fund Denmark (CyberQ, grant agreement no. 3200-00035B).

\bmsection{Data availability} Data underlying the results presented in this paper are available from the authors upon reasonable request.

\smallskip

\bmsection{Disclosures} The authors declare no conflicts of interest.


\bigskip

\end{backmatter}

\bibliography{sample}

\begin{thebibliography}{10}
\newcommand{\enquote}[1]{``#1''}

\bibitem{usenko2026continuous}
V.~C. Usenko, A.~Ac{\'\i}n, R.~All{\'e}aume, \emph{et~al.}, \enquote{Continuous-variable quantum communication,} {\protect\JournalTitle{Reviews of Modern Physics}} \textbf{98}, 015003 (2026).

\bibitem{cain2026shor}
M.~Cain, Q.~Xu, R.~King, \emph{et~al.}, \enquote{Shor's algorithm is possible with as few as 10,000 reconfigurable atomic qubits,} {\protect\JournalTitle{arXiv preprint arXiv:2603.28627}}  (2026).

\bibitem{hajomer2024continuous}
A.~A. Hajomer, I.~Derkach, R.~Filip, \emph{et~al.}, \enquote{Continuous-variable quantum passive optical network,} {\protect\JournalTitle{Light: Science \& Applications}} \textbf{13}, 291 (2024).

\bibitem{qi2024experimental}
D.~Qi, X.~Wang, Z.~Li \emph{et~al.}, \enquote{Experimental demonstration of a quantum downstream access network in continuous variable quantum key distribution with a local local oscillator,} {\protect\JournalTitle{Photonics Research}} \textbf{12}, 1262--1273 (2024).

\bibitem{bian2023high}
Y.~Bian, Y.-C. Zhang, C.~Zhou, \emph{et~al.}, \enquote{High-rate point-to-multipoint quantum key distribution using coherent states,} {\protect\JournalTitle{arXiv preprint arXiv:2302.02391}}  (2023).

\bibitem{liu2024integrated}
S.~Liu, Y.~Tian, Y.~Zhang, \emph{et~al.}, \enquote{Integrated quantum communication network and vibration sensing in optical fibers,} {\protect\JournalTitle{Optica}} \textbf{11} (2024).

\bibitem{Huang2016LongDistance}
D.~Huang, P.~Huang, D.~Lin, and G.~Zeng, \enquote{Long-distance continuous-variable quantum key distribution by controlling excess noise,} {\protect\JournalTitle{Scientific Reports}} \textbf{6}, 19201 (2016).

\bibitem{HuangLin2015_1Mbps}
D.~Huang, D.~Lin, C.~Wang, \emph{et~al.}, \enquote{Continuous-variable quantum key distribution with 1~{Mbps} secure key rate,} {\protect\JournalTitle{Optics Express}} \textbf{23}, 17511--17519 (2015).

\bibitem{Zhang2020LongDistance}
Y.-C. Zhang, Z.~Chen, S.~Pirandola, \emph{et~al.}, \enquote{Long-distance continuous-variable quantum key distribution over 202.81~km of fiber,} {\protect\JournalTitle{Physical Review Letters}} \textbf{125}, 010502 (2020).

\bibitem{huang2015high}
D.~Huang, P.~Huang, D.~Lin, \emph{et~al.}, \enquote{High-speed continuous-variable quantum key distribution without sending a local oscillator,} {\protect\JournalTitle{Optics Letters}} \textbf{40}, 3695--3698 (2015).

\bibitem{leverrier2010cvqkd}
A.~Leverrier and P.~Grangier, \enquote{Continuous-variable {Q}uantum {K}ey {D}istribution protocols with a discrete modulation,}  (2010). ArXiv:1002.4083 [quant-ph].

\bibitem{hajomer2025coexistence}
A.~Hajomer, I.~Derkach, V.~Usenko, \emph{et~al.}, \enquote{Coexistence of continuous-variable quantum key distribution and classical data over 120-km fiber,} {\protect\JournalTitle{arXiv preprint arXiv:2502.17388}}  (2025).

\bibitem{Jouguet2012Analysis}
P.~Jouguet, S.~Kunz-Jacques, E.~Diamanti, and A.~Leverrier, \enquote{Analysis of imperfections in practical continuous-variable quantum key distribution,} {\protect\JournalTitle{Phys. Rev. A}} \textbf{86}, 032309 (2012).

\bibitem{qi2010feasibility}
B.~Qi, W.~Zhu, L.~Qian, and H.-K. Lo, \enquote{Feasibility of quantum key distribution through a dense wavelength division multiplexing network,} {\protect\JournalTitle{New Journal of Physics}} \textbf{12}, 103042 (2010).

\bibitem{jouguet2013experimental}
P.~Jouguet, S.~Kunz-Jacques, A.~Leverrier, \emph{et~al.}, \enquote{Experimental demonstration of long-distance continuous-variable quantum key distribution,} {\protect\JournalTitle{Nature photonics}} \textbf{7}, 378--381 (2013).

\bibitem{kumar2015coexistence}
R.~Kumar, H.~Qin, and R.~All{\'e}aume, \enquote{Coexistence of continuous variable qkd with intense dwdm classical channels,} {\protect\JournalTitle{New Journal of Physics}} \textbf{17}, 043027 (2015).

\bibitem{jain2022Practical}
N.~Jain, H.-M. Chin, H.~Mani, \emph{et~al.}, \enquote{Practical continuous-variable quantum key distribution with composable security,} {\protect\JournalTitle{Nature communications}} \textbf{13}, 4740 (2022).

\bibitem{chin2022digital}
H.-M. Chin, N.~Jain, U.~L. Andersen, \emph{et~al.}, \enquote{Digital synchronization for continuous-variable quantum key distribution,} {\protect\JournalTitle{Quantum Science \& Technology}} \textbf{7}, 045006 (2022).

\bibitem{kanitschar2023finite}
F.~Kanitschar, I.~George, J.~Lin, \emph{et~al.}, \enquote{Finite-size security for discrete-modulated continuous-variable quantum key distribution protocols,} {\protect\JournalTitle{PRX Quantum}} \textbf{4}, 040306 (2023).

\bibitem{hajomer2025experimental_QPSK}
A.~A. Hajomer, F.~Kanitschar, N.~Jain, \emph{et~al.}, \enquote{Experimental composable key distribution using discrete-modulated continuous variable quantum cryptography,} {\protect\JournalTitle{Light: Science \& Applications}} \textbf{14}, 255 (2025).

\bibitem{townsend1997quantum}
P.~D. Townsend, \enquote{Quantum cryptography on multiuser optical fibre networks,} {\protect\JournalTitle{Nature}} \textbf{385}, 47--49 (1997).

\bibitem{frohlich2013quantum}
B.~Fr{\"o}hlich, J.~F. Dynes, M.~Lucamarini, \emph{et~al.}, \enquote{A quantum access network,} {\protect\JournalTitle{Nature}} \textbf{501}, 69--72 (2013).

\bibitem{bian2026approaching}
Y.~Bian, Y.~Zhang, S.~Yu, \emph{et~al.}, \enquote{Approaching the key rate limit in continuous-variable quantum key distribution network,} {\protect\JournalTitle{Physical Review Letters}} \textbf{136}, 080801 (2026).

\bibitem{huang2021realizing}
Y.~Huang, T.~Shen, X.~Wang, \emph{et~al.}, \enquote{Realizing a downstream-access network using continuous-variable quantum key distribution,} {\protect\JournalTitle{Physical Review Applied}} \textbf{16}, 064051 (2021).

\bibitem{pan2025high}
Y.~Pan, Y.~Bian, Y.~Li, \emph{et~al.}, \enquote{High-rate 16-node quantum access network based on a passive optical network,} {\protect\JournalTitle{Optica}} \textbf{12}, 953--960 (2025).

\bibitem{oruganti2025multiuser}
A.~n. Oruganti, \enquote{Multiuser quantum key distribution using quotient graph states derived from continuous-variable dual-rail cluster states,} {\protect\JournalTitle{Physical Review Applied}} \textbf{24}, 054049 (2025).

\bibitem{takeoka2017unconstrained}
M.~Takeoka, K.~P. Seshadreesan, and M.~M. Wilde, \enquote{Unconstrained capacities of quantum key distribution and entanglement distillation for pure-loss bosonic broadcast channels,} {\protect\JournalTitle{Physical review letters}} \textbf{119}, 150501 (2017).

\bibitem{weedbrook2012gaussian}
C.~Weedbrook, S.~Pirandola, R.~Garc{\'\i}a-Patr{\'o}n, \emph{et~al.}, \enquote{Gaussian quantum information,} {\protect\JournalTitle{Reviews of modern physics}} \textbf{84}, 621--669 (2012).

\bibitem{leverrier2010finitesize}
A.~Leverrier, F.~Grosshans, and P.~Grangier, \enquote{Finite-size analysis of a continuous-variable quantum key distribution,} {\protect\JournalTitle{Phys. Rev. A}} \textbf{81}, 062343 (2010).

\bibitem{chinMachine2021}
H.-M. Chin, N.~Jain, D.~Zibar, \emph{et~al.}, \enquote{Machine learning aided carrier recovery in continuous-variable quantum key distribution,} {\protect\JournalTitle{npj Quantum Inf}} \textbf{7}, 20 (2021).

\bibitem{hajomerLongdistance2024}
A.~A.~E. Hajomer, I.~Derkach, N.~Jain, \emph{et~al.}, \enquote{Long-distance continuous-variable quantum key distribution over 100-km fiber with local local oscillator,} {\protect\JournalTitle{Sci. Adv.}} \textbf{10}, eadi9474 (2024).

\end{thebibliography}

\bibliographyfullrefs{sample}
\appendix
\section{Joint key rate decomposition}

\noindent
\begin{minipage}{\textwidth}
    Decomposition of the joint key rate into individual user contributions for all user orders. Each row corresponds to a different conditioning order used in the chain-rule decomposition of the joint key rate $K(A:B_1,\dots,B_4)$. The entries $K_1,K_2,K_3,K_4$ denote the individual user contributions obtained under that decomposition. While the allocation of key among users depends on the chosen order, the total sum $\sum_{i=1}^4 K_i$ is invariant and equals the joint key rate, $0.197827$ bits per channel use for $\beta=0.95$, for all orders.    
    \centering
    \vspace{1em} 
    \begin{tabular}{c c c c c}
        \toprule
        Order & $K_1$ & $K_2$ & $K_3$ & $K_4$ \\
        \midrule
        $(B_1,B_2,B_3,B_4)$ & 0.05962534 & 0.06450573 & 0.04096908 & 0.02890624 \\
        $(B_1,B_2,B_4,B_3)$ & 0.05962534 & 0.06450573 & 0.02878900 & 0.04108632 \\
        $(B_1,B_3,B_2,B_4)$ & 0.05962534 & 0.04094235 & 0.06453246 & 0.02890624 \\
        $(B_1,B_3,B_4,B_2)$ & 0.05962534 & 0.04094235 & 0.02882100 & 0.06461770 \\
        $(B_1,B_4,B_2,B_3)$ & 0.05962534 & 0.02876647 & 0.06452826 & 0.04108632 \\
        $(B_1,B_4,B_3,B_2)$ & 0.05962534 & 0.02876647 & 0.04099688 & 0.06461770 \\
        $(B_2,B_1,B_3,B_4)$ & 0.06445030 & 0.05968077 & 0.04096908 & 0.02890624 \\
        $(B_2,B_1,B_4,B_3)$ & 0.06445030 & 0.05968077 & 0.02878900 & 0.04108632 \\
        $(B_2,B_3,B_1,B_4)$ & 0.06445030 & 0.04091211 & 0.05973774 & 0.02890624 \\
        $(B_2,B_3,B_4,B_1)$ & 0.06445030 & 0.04091211 & 0.02879570 & 0.05984828 \\
        $(B_2,B_4,B_1,B_3)$ & 0.06445030 & 0.02873969 & 0.05973008 & 0.04108632 \\
        $(B_2,B_4,B_3,B_1)$ & 0.06445030 & 0.02873969 & 0.04096812 & 0.05984828 \\
        $(B_3,B_1,B_2,B_4)$ & 0.04083713 & 0.05973056 & 0.06453246 & 0.02890624 \\
        $(B_3,B_1,B_4,B_2)$ & 0.04083713 & 0.05973056 & 0.02882100 & 0.06461770 \\
        $(B_3,B_2,B_1,B_4)$ & 0.04083713 & 0.06452528 & 0.05973774 & 0.02890624 \\
        $(B_3,B_2,B_4,B_1)$ & 0.04083713 & 0.06452528 & 0.02879570 & 0.05984828 \\
        $(B_3,B_4,B_1,B_2)$ & 0.04083713 & 0.02877716 & 0.05977440 & 0.06461770 \\
        $(B_3,B_4,B_2,B_1)$ & 0.04083713 & 0.02877716 & 0.06454382 & 0.05984828 \\
        $(B_4,B_1,B_2,B_3)$ & 0.02866598 & 0.05972584 & 0.06452826 & 0.04108632 \\
        $(B_4,B_1,B_3,B_2)$ & 0.02866598 & 0.05972584 & 0.04099688 & 0.06461770 \\
        $(B_4,B_2,B_1,B_3)$ & 0.02866598 & 0.06452401 & 0.05973008 & 0.04108632 \\
        $(B_4,B_2,B_3,B_1)$ & 0.02866598 & 0.06452401 & 0.04096812 & 0.05984828 \\
        $(B_4,B_3,B_1,B_2)$ & 0.02866598 & 0.04094832 & 0.05977440 & 0.06461770 \\
        $(B_4,B_3,B_2,B_1)$ & 0.02866598 & 0.04094832 & 0.06454382 & 0.05984828 \\
        \bottomrule
    \end{tabular}
\end{minipage}


\end{document}